\documentclass[final,12pt,authoryear]{elsarticle}

\usepackage{amssymb}
\usepackage{amsthm}
\usepackage{amsmath}
\numberwithin{equation}{section}
\usepackage{flexisym}
\usepackage[export]{adjustbox}
\usepackage{subfig}

 \newtheorem{assumption}{Assumption}
\journal{Journal}
\usepackage[authoryear]{natbib}
\usepackage{etex,etoolbox}
\usepackage{amsthm}

\makeatletter
\providecommand{\@fourthoffour}[4]{#4}
\def\fixstatement#1{%
  \AtEndEnvironment{#1}{%
    \xdef\pat@label{\expandafter\expandafter\expandafter
      \@fourthoffour\csname#1\endcsname\space\@currentlabel}}}

\globtoksblk\prooftoks{1000}
\newcounter{proofcount}

\long\def\proofatend#1\endproofatend{%
  \edef\next{\noexpand\begin{proof}[Proof of \pat@label]}%
  \toks\numexpr\prooftoks+\value{proofcount}\relax=\expandafter{\next#1\end{proof}}
  \stepcounter{proofcount}}

\def\printproofs{%
  \count@=\z@
  \loop
    \the\toks\numexpr\prooftoks+\count@\relax
    \ifnum\count@<\value{proofcount}%
    \advance\count@\@ne
  \repeat}
\makeatother

\newtheorem{thm}{Theorem}
\newtheorem{lem}[thm]{Lemma}
\fixstatement{thm}
\fixstatement{lem}

\usepackage{hyperref}
\hypersetup{
     colorlinks   = true,
     citecolor    = blue
}

\begin{document}

\begin{frontmatter}



\title{Optimum Liquidation Problem Associated with \\ the Poisson Cluster Process}


\author{Amirhossein Sadoghi \thanks{Corresponding Author}}
\ead{a.sadoghi@fs.de}
\author{Jan Vecer}
\ead{j.vecer@fs.de}
\address{Frankfurt School of Finance \& Management, Frankfurt am Main}


\begin{abstract}
In this research, we develop a trading strategy for the discrete-time optimal liquidation problem of large order trading with different market microstructures in an illiquid market. In this framework, the flow of orders can be viewed as a point process with stochastic intensity. We model the price impact as a linear function of a self-exciting dynamic process. We formulate the liquidation problem as a discrete-time Markov Decision Processes, where the state process is a Piecewise Deterministic Markov Process (PDMP). The numerical results indicate that an optimal trading strategy is dependent on characteristics of the market microstructure. When no orders above certain value come the optimal solution takes offers in the lower levels of the limit order book in order to prevent not filling of orders and facing final inventory costs.
\end{abstract}

\begin{keyword}
Optimum Liquidation Problem, Discrete order books, Markov-modulated Poisson
process, PDMP, Hawkes processes

\end{keyword}

\footnotetext[1]{Corresponding Author, address: Sonnemannstraße 9-11 60314 Frankfurt am Main, Tel: +49(069)154008-873. We are grateful for the comments of participants of First Berlin-Singapore Workshop on Quantitative Finance \& Financial Risk, Berlin, 2014. Bachelier Finance Society 8th World Congress Brussels, 2014. SIAM Conference on Financial Mathematics \& Engineering, Chicago, 2014. 38th conference Stochastic Processes \& their Applications, 2015, Oxford. \\

\textit{JEL classification}: C60, C61, D4, G1.}

\end{frontmatter}

\newpage


\section{Introduction}
In an illiquid market, due to lack of counterparties and uncertainty about the assets' value, trading of assets by fair value is not guaranteed. In a market like this, a trader faces liquidity restrictions, and behaves differently in comparison to an unconstrained market. Depending on the elasticity of the market, the effect of an increased offer will be compensated by a drop in prices. Effectively, the initial price impact partially is  temporary, and vanishes after the execution of the order depending on the elasticity of the market. Early liquidation causes an unfavorable influence on the stock price, and a late execution has liquidity risk since the stock price can move away from that at the beginning of the period. \\

Most practitioners deal with this dilemma by predicting the trend of stock price and adjusting their order execution speed. While the stock price starts to go down, they  send positive signals to market by decreasing rapidly the size of order; otherwise, they should wait until the end of execution period when the position becomes urgent to liquidate. In such a situation, when they expect the stock price starts to rise, they split orders and execute with offers in deeper levels in the order book to gain an advantage from increasing stock prices. Liquidation of large block orders has an impact on market resilience and market depth. Lack of transparency in a market and splitting trade share help larger traders to avoid reorganization by other participants, who can move against them.\\

In this research, we determine the optimal liquidation strategy of a single trader, who wants to liquidate a large portfolio in an illiquid market. We formulate the liquidation problem as a multi-stopping problem with Poisson arrival patterns of order in a discrete-time model. We set up a stochastic control framework to maximize the expected revenue of trading in the entire execution time taking into consideration the liquidity restriction of trading a large order. The assumption of the discrete-time framework is not far from the classical trading algorithms. In most continuous liquidation problem setups, it is assumed that supply and demand are in balance; this is only possible in a liquid market. We show how the current state of the arrival rates of limit orders in the order book can be used to compute price impact in terms of the conditional distribution of price changes.\\

The optimal liquidation problem can be defined as an optimal control problem to determine optimal strategies for trading stock portfolios by minimizing the entire cost function. These strategies depend on the state of the market as well as price and size of stocks available  during the execution of the limit orders. To find an optimal solution, we need to consider a trade-off between liquidity risk and changing the stock price. The former resulting in in slow order execution and latter caused by exogenous events or a rapid liquidation. Optimal liquidation problems seeks optimal strategies for trading stock with regards to both liquidity risks and non-execution risks. These risks are attributable to liquidity restrictions and time delay of execution orders.\\

The study of the optimal execution problem dates back to 1990's, and mainly focuses on the discrete-time models that optimal strategies are determined as optimal liquidation rates per unit time. \cite{bertsimas1998optimal} studied the optimal liquidation of a large block of shares with linear permanent price impact in a fixed time horizon. \cite{almgren2001optimal} as one of the most cited optimal execution models, used a diffusion price process in a continuous-time trading space. They constructed an efficient frontier of execution strategies via mean-variance analysis of expected costs of liquidation and divided the price impact into temporary and/or linear permanent impact. From a traditional view point on this problem (e.g. \citep{alfonsi2010}, and \citep{alfonsi2010optimal}), the optimal liquidation depends on existence of a sufficiently large limit order in the limit order book, and price impact is a function of the shape and depth of the limit order book.\\

\cite{bayraktar2014liquidation} formulated the liquidation problem as a Hamilton--Jacobi--Bellman equation association with the depth function of the limit order book. A recent study by \cite{horst2014cross} addressed the problem of optimal trading in illiquid markets. They studied a trading algorithm with a two-sided limit order market as well as market orders in a dark pool by controlling the bid-ask spread. Their algorithm determined the optimal time of crossing the bid-ask spread which is a primary problem of algorithmic trading. Under the same market condition, \cite{ horst2011derivatives} studied optimal trading problems in an option market. In their mathematical framework, risk-neutral and risk averse investors hold European contingent claims and the price evolution of the underlying is affected by agents' trading. They proved the existence and uniqueness of equilibrium for a number of interacting players of the market.\\

An optimal stopping problem is a link between the control theory and market microstructure. This problem has been studied as a single or multi-stopping problem in the classical house selling problems or best choice problems by using homogeneous Poisson processes. Single stopping time problem governed with Poisson Process was formulated as a best choice problem in the late 1950's by \cite{lindley1961}. The optimal k-stopping problem with finite and infinite time horizon was presented with the complete solution by \cite{peskir2006optimal} in a Bayesian formulation. A similar problem, which has been considered in the market microstructure literature by \citep{Garman1976}, studied a trading problem of a market maker who maximizes her profit by assuming order arrival rates depending on the price dynamics governed by the Poisson process. \cite{smith2003} simulated the buy and sell order with thr Poisson process in which the arrival rates are independent of the state of the order book. Recent study by \citep{cont2010} used Laplace transforms to analyze the behavior of the order-book with Poisson arrivals of buy and sell orders.\\

The homogeneous Poisson Process of sell and buy orders arrival rate was the main assumption of past studies. \cite{Garman1976} explained conditions necessary and sufficient for the order arrival patterns to be modeled by a homogeneous Poisson processor. In this framework, none of the agents' trading can dominate other agents or they place a large number of limit orders in a finite time. Nevertheless, most of these assumptions are violated in the high-frequency trading. Empirical studies of high-frequency data show that there are significant cross-correlation patterns of the arrival rate of similar limit buy or sell orders and significant autocorrelation in durations of events, (see: \cite{cont2011}).\\

Some empirical studies like \cite{Bacry2013} showed that high frequency data can be modeled by Hawkes processes, introduced by \cite{HAWKES1971}. The irregularity properties of high frequency financial data can be explained by self-exciting and mutually-exciting properties of the Hawkes processes. 
\cite{engle2003trades} modeled trade and quote arrivals with bivariate point processes.\\

In the illiquid market, the disparity between supply and demand causes illiquidity, which means the executing larger positions need longer time. We formulate the liquidation problem as a discrete-time Markov Decision Processes (MDP) where the state process is a Piecewise Deterministic Markov Process (PDMP), which is a member of right continuous Markov Process family introduced by \cite{davis1984piecewise}. By applying a MDP approach, we decompose the liquidation problem as a continuous-time stochastic control problem into multiplies deterministic optimal control problems and construct an approximation of value function with a quantization method proposed by \cite{bally2003quantization}. In our numerical simulation, we compare the performance of our algorithm with regard to various market characteristics and price impact functions.\\

The numerical results show that the trading strategy is dependent on the characteristics of market microstructure and dynamics of incoming orders. In the case of favorite offers are not coming, our algorithm will reduce the speed of trading and allows the trader to go deeper into limit order book to avoid not filling of orders and facing an ultimate inventory penalty. The simulated results indicate that higher probability of same types of orders' occurrences (self-exciting property of the point process), creates more profitable trading opportunities.\\

The remainder of this paper is organized as follows. Section \ref{Statistics_Model_OrderBook} describes the statistical model of order book. Section \ref{MarkeModelSetup} explains the market model setup and present the problem statement. Section \ref{Stochastic_Intensity} presents the stochastic dynamic of the intensity of the order arrival process. In section \ref{Price_Impact}, we turn to model price impact. Section \ref{Solving_model} describes the procedure of solution with using discrete-time Markov Process. In section \ref{Numerical}, we explain the numerical method for optimal stopping time and simulate with different micromarket structures. Section \ref{summarizes} summarizes the results, and concludes the paper with further remarks.

\section{Statistics Model of Limit Order Book}
\label{Statistics_Model_OrderBook}
To understand the behavior of market participants, we need to analyze the stochastic fluctuation of stock price and the reaction of players in the market. These fluctuations can be explained by a sequence of equilibria of demand and supply in the market. In order-driven markets, buy and sell orders arrive at different time points and wait in the LOB (Limit Order Book) to be traded. The limit order book contains information about orders like quantity, price and type of orders. 
This information can be used to reduce the complexity of the relation between price fluctuations and limit order book dynamic. It also help to predict different market's quantities conditional on the current state of the order book. On other hand, this Information might be used against counter-parties to take advantage of price movement. The idea of the dark pool to reduce the information linkage and adverse price risk.\\

\cite{cont2010} shows that this information contains short-term price movements and might change quickly during the trading period. In illiquid markets, the relation between limit order book dynamics and the price behavior shows that the less liquid assets wait for a longer time to be executed. Therefore queue type models can be applied to analyze the orders arrival in the limit order book.  \cite{cont2010} developed a tractable stochastic model of limit order markets to capture the main statistical features of limit order books.\\

The statistical characteristics of the limit order book are well-documented by \cite{smith2003}. \cite{ Garman1976}, \cite{Bayraktar2007} studied the fluctuation of the orders arrival in the electronic trading market. Some studies explained the unconditional and steady-state distributions of the order book while \cite{cont2010} showed the short-term price movements could be explained by the information on the current state of the limit order book. Their model is based on the dynamics of the best bid and offer queues since the best orders can move the price.\\

Underlying our approach is that the dynamics of limit orders arrivals follow a Poisson pattern of prices depending on rates of trading. We study the stochastic dynamics of the arrival rate of limit order with the help of order statistics.\\
 
\subsection*{Distribution of a limit order book}
We assume in a financial market, the price processes $(S_t^j)_{(j=1 \cdots d)}$ of limit orders  satisfy the following stochastic dynamic in a finite discrete-time horizon $t \in \tau = \{ 1<T_1 \dots < T_m<n \}$:
\begin{equation}
S^j_t = S^j_0 (\mu_j dt+ dC^j_t),
\end{equation}  
where $C_t = \sum_{n=1}^{N_t} J_n$, and $J_n$ is an independent and identically distributed random variable and $N_t$ counting process.\\ 
Given a vector of price of buy (sell) orders $(S^j_t)_{1 \leq j \leq d}$ on the probability space $(\Omega;\mathcal{F};\mathbb{P})$,      
the orders are sorted into a vector $(S^{(1)}_t,S^{(2)}_t,\cdots, S^{(j)}_t)$ satisfy:\\
\begin{center}
$(S^{(1)}_t \leq S^{(2)}_t,\cdots, \leq S^{(j)}_t)$
\end{center}
The vector $(S^{(1)}_t,S^{(2)}_t,\cdots, S^{(j)}_t)$ is a so-called vector of order statistics of the price processes ($S_t^j$).\\

Let $L$ be the number of unexecuted orders at the time point $s$. In the time interval $(s,t)$, the rate of arrival of limit orders is governed by a Poisson process. It is assumed that noisy traders cannot have a strong influence on the market, and no cancellation or strategic cancellation of orders can occur. Hence, this model is a pure birth Poisson process with intensity $\lambda$, and the arrivals of orders in the non-overlapping intervals are independent. \\


In a small time interval $dt$ , the probability that a limit order arrives and sits in the limit order book is:
\[
    \begin{array}{ll}
P(N(dt)=1)= \frac{\lambda dt}{1!} e^{- \lambda dt}= \lambda dt + O(dt)
    \end{array}
\]
Similarly, the probability that no limit order appears in the limit order book in this interval is:
\[
    \begin{array}{ll}
P(N(dt)=0)= e^{- \lambda dt}= 1-\lambda dt + O(dt)\\ 
    \end{array}
\]

\begin{figure}[h]
\begin{center}
   \includegraphics[scale=.25]{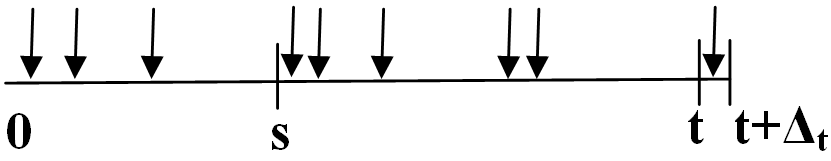} 
      \caption{Orders' arrival  governed by Poisson process}
\end{center}
\end{figure}

\newtheorem{theorem}{Theorem}
\begin{thm}[\textbf{Distribution of the limit order book}]
\label{thm:goldbach}
Denote by $L$ the number of unexecuted orders at time $s$. We assume that in the interval $(0,t)$, the number of orders ($N_t$) is a random variable with the Poisson distribution with mean value $\lambda  t$. Vector $\{S^1_t,\cdots, S^{N_t}_t\}$ represents the prices of buy (sell) orders with distribution $F(S)$ on the probability space $(\Omega;\mathcal{F};\mathbb{P})$. $U$ is the number of unexecuted orders with prices greater than $Y= \max\{S^1_t,\cdots, S^{N_t}_t\}$, then the probability of having $k$ unexecuted orders with prices greater than $Y$ is:\\
       \begin{eqnarray*}
P(U=k|N_s=L) = \frac{e^{-\lambda_Y} \cdot \lambda_Y^k}{k!},
    \end{eqnarray*}
where   
    \begin{eqnarray*}
    \label{dis_order_book}
\lambda_Y=\lambda [1-F(Y)t-\ln(e^{\lambda F(Y)}- \frac{\lambda^{L+1}}{(L+1)!})].
    \end{eqnarray*}
\end{thm}   
(Proof is given in the appendix)\\
With given $k=0$ , equation \ref{dis_order_book} gives the probability of best orders (i.e. none of orders with price $(S^i_t)_{i\in \{1,\cdots,N_t\}}$ has greater price than $Y$), with $k=1$ gives the probability of the second best order,and so on. 

\newtheorem{lemma}{Lemma}
\proofatend
If the number of orders is a random variable with the Poisson distribution and the mean value $\lambda$ in a finite interval of length $t$. We consider the order with maximum price of $N^{(0,t)}$ number of limit orders as:\\
\begin{center}
 $Y= \max\{S^1_t,\cdots, S^{N^{(0,t)}}_t\}$
\end{center}
We define  $F_{N^{(0,t)}}(Y)$
\begin{eqnarray*}
F_{N^{(0,t)}}(Y) &=& P[(S_1 \leq Y ) \cap (S_2 \leq Y) \cap ,\cdots, S_N \leq Y ) ]\\
         &=& F(Y)^{N^{(0,t)}}
\end{eqnarray*}
The generating function with distribution function $F(S_t)$ is
\begin{eqnarray*}
 G_t(F(y))= \mathbb{E}[F(Y)^{N^{(0,t)}}]
\end{eqnarray*}

\begin{eqnarray*}
 G_{t+dt}(F(Y))&=& \mathbb{E}[F(Y)^{N^{(0,t+dt)}}]\\
                 &=& \mathbb{E}[F (Y)^{N^{(0,t)}+N^{(t,t+dt)}}]\\
                 &=& \mathbb{E}[F(Y)^{N^{(0,t)}} \mathbb{E}[F(Y)^{N^{(t,dt)}}]\\
                 &=&  G_t(F(Y)) (1- \lambda dt + \lambda dt .F(Y) )
\end{eqnarray*}

\begin{eqnarray*}
 \frac{G_{t+dt}(F(Y)) - G_t(F(Y))}{dt} = - \lambda (1- F(Y)) G_t(F(Y))
\end{eqnarray*}
\begin{eqnarray*}
\frac{d}{dt} (G_t(F(Y)) = - \lambda (1- F(Y)) G_t(F(Y))
\end{eqnarray*}
\begin{eqnarray*}
 \frac{d}{dt}\ln (G_t(F(Y))  = - \lambda  (1- F(Y))
\end{eqnarray*}

\begin{eqnarray*}
 G_t(F(Y)) = e^{- \lambda t(1- F(Y))}
 \label{equationGT}
\end{eqnarray*}

With assumption of a $L$ number of unexecuted orders at the time point $s, (s<t) $, the generating function for the time interval $(0,s)$ is:

\begin{eqnarray}
G_s(F(y)) &=& \frac{\lambda^0}{0!}e^{-\lambda}(F(Y)^0)+\frac{\lambda^1}{1!}e^{-\lambda}(F(Y)^1)\\
            &+& \cdots + \frac{\lambda^L}{L!}e^{-\lambda}(F(Y)^L)\\
            &=& e^{-\lambda}(\frac{(\lambda F(Y))^0}{0!} + \frac{(\lambda F(Y))^1}{1!} + \cdots + \frac{(\lambda F(Y))^L}{L!})
\end{eqnarray}    
 Use Taylor series with remainder:
   
    \begin{eqnarray}
    \label{eqaGs}
G_s(F(Y)) =e^{- \lambda}(e^{\lambda (F(Y))}-\frac{f(c) \lambda^{L+1}}{(L+1)!})
\end{eqnarray}
For $c\in [0,1]$, $f(c)=F(Y)^{L+1}\approx e^{c \lambda}$. For the sake of simplicity, it is assumed that $c=0$. \\
\begin{eqnarray}
G_t(F(Y))&=& G_{s+(t-s)}\\
           &=& G_s(F(Y))G_{t-s}(F(Y))
\end{eqnarray}
From equations \ref{equationGT} and \ref{eqaGs}, the generating function of time interval $((t-s),t)$:
\begin{eqnarray}
G_{t-s}(F(Y)) &=& \frac{e^{-\lambda (1-F_x(y))}}{e^{- \lambda \ln(e^{\lambda (F(Y))-\frac{\lambda^{L+1}}{(L+1)!}})}}\\
                &=& e^{-\lambda[(1-F_x(y))t-\ln(e^{\lambda F(Y)}-\frac{\lambda^{L+1}}{(L+1)!})]} 
\end{eqnarray}
Consider the number of orders in the coming stopping time is a random variable with Poisson distribution and mean value $\lambda t$. From above generating function, we can define the probability that no order arrival in time interval $(0,t)$ is said to have a Poisson distribution greater than $Y$ :
 \begin{eqnarray*}
P(U=k | N^{(0,s)}=L) = \frac{e^{-\lambda_Y} \cdot \lambda_Y^k}{k!}
    \end{eqnarray*}
    
    \begin{eqnarray*}
\lambda_y=\lambda  [(1-F(Y)t-\ln(e^{\lambda F(Y)}- \frac{\lambda^{L+1}}{(L+1)!})]
    \end{eqnarray*}
where $F(Y)$ is the distribution of the price process and $L$ is the number of unexecuted orders up to time point $s$. Where  $k=0$ , it gives the probability of the best order, $k=1$ is the second the best order, etc.
\endproofatend

\section{Market Model Setup}
\label{MarkeModelSetup}
It is assumed that our financial market consists of an illiquid asset, it can be considered as a risky asset, and a risk-free asset as a numeraire with the interest rate $r$. The market for the risk-free asset is liquid, that means traders can liquidate a large amount of this security without facing costs of price impact.\\

 In a complete financial market with a finite discrete-time horizon  $\tau = \{ 1 < T_1 \dots < T_m <n \}$, the price process $S_t$ is a stochastic process on a complete filtered probability space $(\Omega;\mathcal{F};\mathbb{F};\mathbb{P})$ where $\mathbb{F}$ is the filtration generated by $\{\mathcal{F}_t\}_{t \in \tau}$. This space is bounded by a maturity time $T$.\\
 
The assumption of the finite discrete-time space is contradiction to conventional liquidation problems. However, it is not far away from reality; order arrival patterns of buy and sell orders are not the same. Because of the fluctuation in the stock market, especially in a high-frequency environment, orders might not  appear regularly in the LOB. Some empirical studies show that in a short period of time, the percentage changes of the stock price are not uniformly distributed with same centrality, but price process can be internally steady state processes. Therefore, the model should be solved and interpreted in the discrete time and space. We use these facts and propose a model with the rate of incoming buy orders as a Poisson process in a finite discrete-time and space.\\



\subsection{Trading Boundaries}
 Traders often follow trends of the prices and construct trading boundaries built in on market depth. They submit limit orders at touch with an optimum volume when the stock prices hit one of these trading boundaries. Market depth reflects the information related to the prices of buy and sell orders for the price and depends on trade volume, and minimum price increment known as tick size. It is constantly changing, and reflects the valuable information about the current orders sitting in the limit order book. Knowing this information helps traders to understand hidden patterns of price movements and price impact. Traders can use this potential information, and regulate their orders in response to the net order flow.\\

 \cite{cont2014price} studied the link between price impact and market depth and showed that there is a linear relationship between an orders flow imbalance and price changes in high-frequency markets. \cite{kempf1999market} measured the market depth as a surplus demand amount that is needed for the jumping of one unit price. They also showed that the price is sensitive to change from counterparties' demand. They concluded that there is a non-linear relation between market depth and price impact of the orders flow. \cite{bessembinder1993price} studied price, trading volume, and market depth in futures markets, and showed that the price impact has negative relationships to liquidity, i.e. liquidity provisions decrease during trading activities in an illiquid market.\\

\begin{figure}[h]
\begin{center}
   \includegraphics[scale=.35]{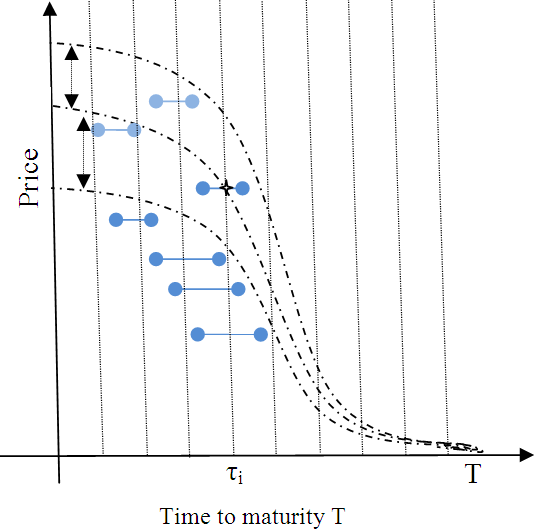} 
      \caption{Schematic model of trading boundaries built on market depth with Poisson arrival patterns}
 \label{schemanticboundry}     
\end{center}
\end{figure}

Market depth gives the overall picture of market conditions and a short term prediction to determine an optimum strategy, e.g. price movement towards selling pressure or buying strength in the short time, especially in illiquid markets. Some platforms apply a number of restrictions for market participants to observe or trade at the best level; we assume all orders are observable and accessible for traders. Market depth can be affected by the transparency of markets in such away that some levels are hidden, and just latest submitted orders are available. Limiting  trading and adjusting minimum price increment, known as the tick size are important mechanisms to improve the efficiency of the market.\\

In illiquid markets, lower price levels of market depth are more attractive than upper levels; these depth levels include orders with significantly larger volumes. Most optimal liquidation methods focus on the best sell and buy price levels since their imbalance can move prices. By cause of the lack of available liquidity, we need to take the lower level of market depth into account. We will show how to explain the dynamic of different levels of market depth with Point processes.\\
 
\subsection{Problem statement}
The dynamics of the price is described by a right-continuous process $(S_t)_{(t \geq 0)}$ changing while the times when the book order process meets boundary conditions. Later on, we will discuss the impact of the current trade on the underlying price by using a particular approach to model price impact based on exogenous factors as well as characteristics of the stock price processes. We will show how to model the dynamics of the intensity of orders arrival  corresponding feedback effects of trading and the state of the market during orders execution period.\\

We consider a discrete-time problem for an investor holding a large volume ($Q^0$) shares of illiquid assets and a risk-free asset. The objective of the shareholder is to maximize her profit with subject to liquidity constraints and a limited execution time.\\
The illiquid asset, also can be considered a risky asset with the following dynamic in a finite discrete-time horizon $ \tau = \{ 0<T_1 \dots < T_m <n \}$:
    \begin{eqnarray*}
dS_t=S^0(\mu_t dt+ \sigma_t w_t) \quad t \in \tau
    \end{eqnarray*}
and a risk-free asset used as a numeraire with the interest rate $r$ with the following dynamic:
    \begin{eqnarray*}
dB_t=B^0 e^{rt}
    \end{eqnarray*}\\
Let $\hat{S}=e^{-rt} S_t$ be a martingale with respect to the  measure $\mathbb{P}$ on a complete filtered probability space$(\Omega;\mathcal{F};\mathbb{P})$.\\
The trading strategy of the trader is characterized by a trading rate $(\gamma_t)_ {t \in \tau}$ . The vector $\boldsymbol{\gamma}$ contains the information on the amount of trading at each time point $t$.\\
The dynamics of the inventory of the investor holding $Q^0$ shares of an illiquid asset with the trading rate $\boldsymbol{\gamma}$ as a control process is given with the following counting process:
    \begin{eqnarray*}
d{q_t}^{\boldsymbol{\gamma}}=-\bigtriangleup \gamma_t d {N_t}^{\boldsymbol{\gamma}}
    \end{eqnarray*}
 where the $\bigtriangleup$ is a fraction of $Q^0$ shares, assumed to be constant at each stopping time i.e. either we fill whole orders or reject offers. This assumption can be questionable, but from the theoretical point of view, it reduces the complexity of the problem. Let $N_t$ be a counting process, and $F_t$ be a cash flow process with dynamics:
       \begin{eqnarray*}
d{F_t}^{\boldsymbol{\gamma}}= \hat{S_t}\bigtriangleup \gamma_t d {N_t}^{\boldsymbol{\gamma}}
    \end{eqnarray*}  
The investor has a finite time to liquidate her risky assets and maximize her wealth:

       \begin{eqnarray*}
\sup_{\boldsymbol{\gamma} \in \Psi(Q^0)}(W_T)
    \end{eqnarray*} 
where $W_T$ is the amount of the cash at the end of time horizon $T$.  We define $\Psi(Q^0)$ as a set of admissible strategies with given initial inventory with $Q^0$ shares to have nonnegative inventory at all times:

 \begin{eqnarray*}
 \mathcal{A}(Q^0) \equiv \{ \boldsymbol{\gamma} :\boldsymbol{\gamma} \mbox{ is a predictable process, and }\\ \mbox{admissible strategy from the initial inventory } Q^0\}
    \end{eqnarray*} 
We shall use $\mathcal{A} \equiv \mathcal{A}(Q^0)$ for the set of admissible strategies $\boldsymbol{\gamma}$.\\

The liquidation problem can be formulated as a multi-stopping problem with discrete time sequences $\{T_1, \cdots, T_n \}$. The trader can liquidate her shares in the $m$ stops $1 \leq T_1 < \cdots < T_m \leq n$ with a discrete limit order book. The goal is to maximize the expected gain at each stopping sequence $T_i$ $(i \leq m)$. The interaction between price impact and price dynamics that makes the execution-cost control a dynamic optimization problem. We apply the Bellman's optimality equation in a recursive format to solve the m-stopping problem.
    \begin{eqnarray*}
\max[\mbox{Expected Revenue}]&=& \max[\mbox{Immediate Exercise}\\
                            &+& \mbox{Continuation} ]
 \end{eqnarray*}
 
The trader holds $Q^0$ number of shares and places a selling order an $k= \bigtriangleup \boldsymbol{\gamma}$ illiquid asset in the market with the trading rate $\boldsymbol{\gamma}$ in the time horizon $T$, the performance criteria with the strategy $\gamma$ is given by:
    \begin{eqnarray*}
    H^{\boldsymbol{\gamma}}(t,Q^0)  &=&  [h^{\boldsymbol{\gamma}}(t,k)+ \mathbb{E}_{t,q} [H^{\boldsymbol{\gamma}}(t+1),Q^0-k]]
    \label{maincerterifunction}
   \end{eqnarray*}
with a boundary condition: $h(T,0)=h(0,Q^0)=0$.\\
At the stopping time, when one of these expected boundaries (see figure \ref{schemanticboundry})is hit by the stock price, we execute some proportion of the inventory via limit order at the LOB with the depth function $h(t,k)$:
  \begin{eqnarray*}
  \label{mainformalprobelm}
h(t,k)&=& \max[(\mathcal{S}_{(t,k)}-\mathbb{E}S_{(t,k)},0)]\\
       &=&(\mathcal{S}_{(t,k)}-\mathbb{E}X_{(t,k)})^+
      \end{eqnarray*}
      where:
           \begin{eqnarray*}
\mathcal{S}_{(t,k)}=\sum_{i=n-k+1}^n S^{(i)}_{t} \mathbf{1}_{(i\leq k)}
  \end{eqnarray*}
    \begin{eqnarray*}
  \mathbb{E}\mathcal{S}_{(t,k)}=\sum_{i=n-k+1}^n P(i\leq k) S^{(i)}_{t}
    \end{eqnarray*}
    
Let $\mathbb{E}\mathcal{S}_{(t,k)}$ be the expected boundaries, constructed from the set of best limit orders in the LOB by using the distribution of the order book at the stopping time $t$.\\

The goal of the investor is to maximize her wealth at end of the time period $T$ associated to this dynamic problem, the value function can be defined by:    
    \begin{eqnarray*}
    V(t,q)  &=& \sup_{\gamma \in \mathcal{A}}\mathbb{E}_{t,q} [H^{\boldsymbol{\gamma}}(t,q)], \quad  t \in \tau , \quad q \in [0,Q^0].
   \end{eqnarray*}
The terminal wealth value function of the investor who maximizes her wealth at the end of time period $T$ and give discounted price process $\hat{S}$, and inventory $Q^0$, associated to optimal trading rate $\boldsymbol{\gamma}$, can be expressed with the following lemma:\\

\begin{lem}
Let $V(T,Q^0)$ be the continuation value that is obtained from optimal trading of $Q^0$ shares until the end of period $T$ and the intensity process $\lambda_t$ be defined as a rate of arrival of limit orders. The expected revenue from execution of limit orders with arrival patterns with Poisson distribution is equal to:
        \begin{eqnarray*}
V(T,Q^0)= \sup_{\boldsymbol{\gamma} \in \mathcal{A}} \mathbb{E}_{t,q}[\int_{0}^{T \prime}e^{-r \tau} \bigtriangleup \gamma_t S_t  \lambda_t dt ],
    \end{eqnarray*}
where $T \textprime = T \wedge \inf \{ t \geq 0 : Q^0-q_t =0\}$ is trading time and trading process $\gamma_t$ is a control process, which has influence over the cash process, the inventory process, and the dynamics of prices.
\end{lem} 
(Proof is given in the appendix)\\
           
\proofatend
It is assumed the limit orders arrival rate is a point process with intensity rate $\lambda_t$ and we liquidate $\bigtriangleup \gamma_t$ at each stopping time: $\tau= \{0< T_1 < T_2 <\dots < T_n \leq T\}$
             \begin{eqnarray*}
V(T,Q^0)&=& \sup_{\boldsymbol{\gamma} \in \mathcal{A}}\mathbb{E}_{t,q}[V(T,Q^0)] \\
      &=&  \sup_{\boldsymbol{\gamma} \in \mathcal{A}} \mathbb{E}_{t,q}[\sum_{i=1}^{n}e^{-r T_{i}} \bigtriangleup \gamma_t S_t  1_{(T_{i} \leq T)}]\\
   (\mbox{when n} \to \infty) (t \in \tau)  &=&  \sup_{\boldsymbol{\gamma} \in \mathcal{A}} \mathbb{E}_{t,q}[\int_{0}^{T \prime}e^{-r t} \bigtriangleup  \gamma_t S_t dN_t ]\\
        &=&  \sup_{\boldsymbol{\gamma} \in \mathcal{A}} \mathbb{E}_{t,q}[\int_{0}^{T \prime}e^{-r t} \bigtriangleup \gamma_t S_t  \lambda_t dt ]\\
        \label{mainproblem}
    \end{eqnarray*}    
\endproofatend

\section{Modeling Stochastic Intensity}
\label{Stochastic_Intensity}
The rate of arrival of limit orders depends on the price and size of orders; the cheaper orders will be remained for a shorter time on the limit order book. It is empirically shown that the distribution of the price is not constant and depends on the current state of the limit order book. The existence of significant autocorrelation of price movement and correlations across time periods rejects the natural assumption of a constant intensity of order arrivals rate (see. e.g., : \cite{cont2011}).\\

In an illiquid market, bid and ask orders do not arrive consistently, and counterparties do not meet their demands regularly. These irregular proprieties of a high-frequency environment lead to applying point process to model time series of bid or ask price movements. \cite{avellaneda2008high} and \cite{Garman1976} proposed models which order arrivals are governed by a point process with constant intensity. There are several conditions in previous trading models that do not hold our setting. Firstly, a small number of traders cannot dominate the market with a large-scaled orders. Second condition is the submitting of orders are independent and mainly we have a assumption of efficiency of the market. These conditions are essential to have a constant intensity of order arrivals. However, the structure of the market is dynamic, and high-frequency traders dominate over seventy percents of the market; therefore, none of these above conditions can be satisfied.\\

Hawkes process as a point process introduced by \cite{HAWKES1971}, was initially is applied to model earthquake occurrences. Some recent empirical studies show that Hawkes process can fit with high-frequency data to explain its irregularity properties based on the positive and negative jumping behavior of the asset prices. \cite{cartea2014buy} used the Hawkes processes to express the dynamics of market orders, the limit order book as well as effects of adverse selection. This process is a general form of a standard point process, the intensity is conditional on the recent history, are increase the rate of arrival the same type event (Self-exciting property), and it captures the impact of arrivals of orders on other type of orders (Mutually-exciting property).\\

The concept of a point process is fundamental to the stochastic process. Before we explain the dynamic of the Hawkes process, we state the following formal definition of a point process, a counting process and an intensity process:
\newtheorem{mydef}{Definition}
\begin{mydef}
\textbf{Point Process} : \\
Let $t_i$ ($i\in \mathbb{N}$) be a sequence of non-negative random variables, which is measurable on the probability space  $(\Omega;\mathcal{F};\mathbb{P})$, in such a way that $\forall i \in \mathbb{N}, t_i < t_{i+1}$, is defined as a point process on $\mathbb{R}$.  
\end{mydef}
\begin{mydef}
\textbf{Counting Process} : \\
The right-continuous process $ N(t) = \sum_{i\in N} \mathbf{1}_{t_i\leq t}$ ,with given a point process $t_i$ ($i\in \mathbb{N}$), is a so-called counting process if it measures the number of discrete events up and including the time point $t$.
\end{mydef}
\begin{mydef}
\textbf{Intensity Process} : \\
With given $N_t$ as a point process adapted to a filtration $\mathcal{F}$, the intensity process $\lambda$ as a left-continuous process is defined by:
\begin{eqnarray}
\lambda(t|\mathcal{F}_t) &=& \lim_{\Delta t \to 0} \mathbb{E}[\frac{N_{t+\Delta t}-N_t }{\Delta t}|\mathcal{F}_t ]\\
                        &=& \lim_{\Delta t  \rightarrow 0} \mathbb{P}\frac{1}{\Delta t}[N_{t+\Delta t} - N_t |\mathcal{F}_t ]\
\end{eqnarray}
\end{mydef}
To be more precise, a intensity process $\lambda_t$ is determined by a counting process $N_t$ with the following probabilities:

\begin{eqnarray}
\mathbb{P}(N_{t+\Delta t} - N_t = 1) &=& \lambda \Delta t + O(\Delta t)\\
\mathbb{P}(N_{t+\Delta t} - N_t = 0) &=& 1- \lambda \Delta t + O(\Delta t)\\
\mathbb{P}(N_{t+\Delta t} - N_t > 1) &=&  O(\Delta t)
\end{eqnarray}

Homogeneous Poisson process is  a so-called intensity process that is independent of the probability of the occurrence in the small interval $\Delta t$ and a filtration $\mathcal{F}_t$.\\
A general form of a linear self-exciting process can be expressed:

\begin{eqnarray}
\lambda_t = \lambda_0 + \int_{-\infty}^{t} \sigma f(t-s) dN_s
\end{eqnarray}

where $\lambda_0$ is a deterministic long run "base" intensity,which is assumed to be constant. The function $f(t)$ expresses the impact of the past events on the current intensity process, and the parameter $\sigma$ explains the magnitude of self-exciting and the strength of an incentive to generate the same event.\\

As a result of converting the integrated intensity into independent components via exponentially distributed variables, we can apply most of analytical methods to analyze these statistical random variables. One can calibrate the parameters of the Hawkes process with parametric estimation methods like maximum likelihood estimation (\cite{raey1979}), or with non-parametric estimators like Expectation-Maximization (EM) algorithm (\cite{ raey2012}). The goodness of the fit of these models also can be examined with conventional statistical tests.\\

\begin{figure}[h]
\begin{center}
   \includegraphics[scale=.35]{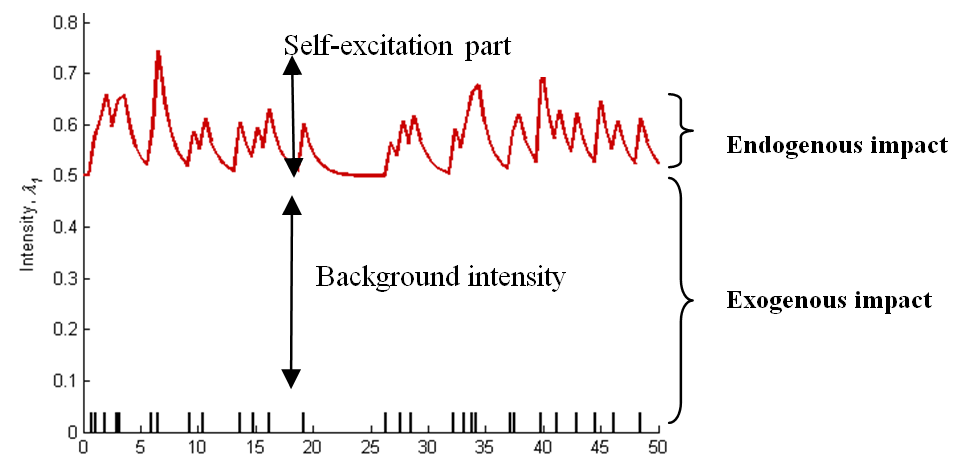} 
      \caption{Hawkes Process with 31 events}
\end{center}
\end{figure}


\section{Price Impact Model}
\label{Price_Impact}
Estimating and modeling price impacts is a crucial research in the market microstructure literature. It can be expressed as a relationship between trading activities and price movements. Monitoring and controlling the impact of trading are the main part of algorithm trading. To minimize the market impact, traders split their orders into smaller chunks based on the current liquidity in the limit order book. Price impact might be dependent on exogenous factors like trade rate and some endogenous factors such as liquidity and volatility.  It is empirically observed that there is a changing of volatility of prices on trading activities. \cite{kyle1985continuous} proposed a simple model for the evolution of market prices and price movement. In his model, noise traders and inform traders submit orders and in the each step, and a market maker executes the orders. The price is adjusted to a linear relationship between the trade size and a proxy for market liquidity. As a consequence of trading, price moves permanently, information affect the price for a long time, and price changes are strongly autocorrelated. The most recent literature on the market microstructure shows that for liquid markets price impact cannot be permanent. In highly liquid markets, outstanding shares are small and need less time to be liquidated. \\

In illiquid markets, the temporary price impact governs the enduring impact, and after a considerable number of trades, the price movement shows a high resilience. Empirical studies show that an elastic market can be converted to a plastic (inelastic) market as a consequence of lacking counterparties, such that high volume  trading have a long-lasting impact. In this market traders face difficulty to find counterparties at the particular time, and they should wait for a longer time to execute the orders or else cooperate with counterparties. The effect of price impact corresponding to the liquidation strategy can be significantly large for substantial risk aversion traders who liquidate shares at a fast rate. Modeling of price impact on illiquid markets is not so well studied; we review some different generations of price impact modeling in literature.\\

We can loosely classify price impact models based on their effects and how long they have influence over the dynamics of the prices in four categories: The first class is a permanent price impact. As a consequence of a significant discrepancy between supply and demand in the market and a spread of bulk trading information, the dynamic of price have stable shifts. \cite{kyle1985continuous} proposed a basic microstructure model to analyze the price impact. Permanent impact was one the component of \cite{bertsimas1998optimal} and \cite{almgren2001optimal} price model. Empirical researches show that in a liquid market trading activities cannot alter the dynamics of the price for a long time. Conversely, this component can be one of the basic building blocks of price impact models in the illiquid market.\\

The second tier of the price impact model is related to modeling the temporary impacts that just have an effect on the current orders for a short time and not for the entire trading time. This component cannot alter the dynamics of price for a long period and just has an impact on the immediate execution of the trades. The transient impact is a third class of price impact; this component can be significant for a finite period, and it eventually vanish [\cite{gatheral2010}]. \cite{alfonsi2008constrained}; \cite{ Shreve2011}) modeled price impact by considering this component.\\

The last class of the price impact modeling is to control the rate of arrival of limit orders via the trading rate process.  \cite{ alfonsi2014dynamic} introduced a mixed market impact Poisson model to analyze a temporary shift of dynamics of the rate of order arrival. This model used the advantage of the self-exciting property of the Hawkes process to change the direction of trading in the same or opposite direction of order arrivals.  \cite{bayraktar2014liquidation} and \cite{gueant2012optimal} controlled the intensity of limit orders for the liquidation problem in a risk-neutral and risk averse model, respectively.\\ 

In this paper, we introduce a stochastic intensity process to measure the price impact of order executions. This model is associated with a counting process using the mutual-exciting property of the Hawkes process. In illiquid markets, the imbalance between supply and demand causes illiquidity and the execution of a larger position need a longer time and affect on the arrival of the new orders or stay longer in the limit order book. Also, we explained the coming pattern of orders governed by a stochastic point process. Therefore, the intensity of order arrival is influenced by the arrival time of orders and the orders' value, and indirectly the price, will be affected by trading.\\

\begin{figure}[h]
\begin{center}
   \includegraphics[scale=.3]{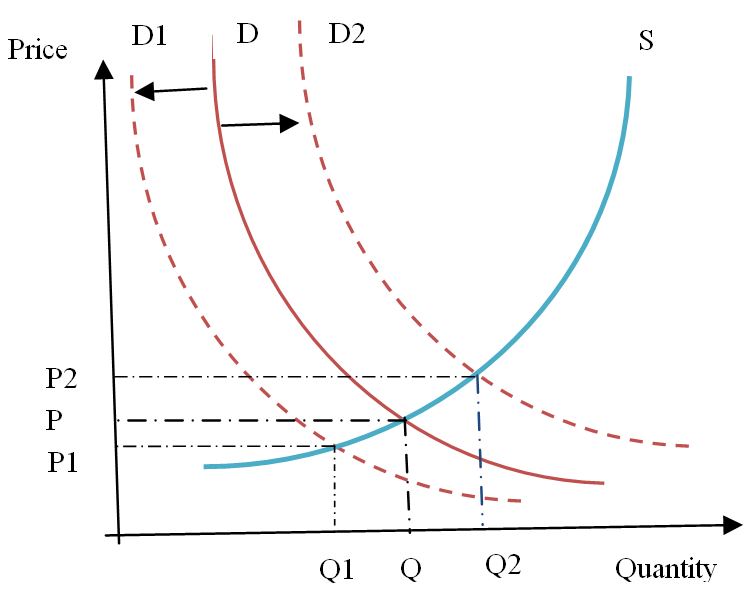} 
      \caption{Changes in equilibrium price as a result of increasing and decreasing of demands}
 \label{supplydemand}     
\end{center}
\end{figure}

According to fundamental concepts of economics, in equilibrium, the relationship between supply and demand determines the price that traders are willing to take positions during a specified period (see: figure \ref{supplydemand}). The price movement occurs when a change in demand is caused by a change in a number of market orders. This change in the price equilibrium is sensitive to shifts in both prices, and quantity demanded, which is called the "price elasticity of demand."

We model this impact with close form solutions for the dynamics of price impact. Let function $f(\gamma_t)$ be a general form of the impact market, conditional on the state of liquidity of the market. Similar to \cite{kyle1985continuous} model, we use parameter $\alpha$, as inverse liquidity of the market, to measure price impact. Function $\Gamma(t-s)$ is the  decay of price impact function, which is independent of the state of the market. This function can be shown by the exponential or power law decay of price impact.\\

Different types of price impact  and decay functions have been used in the algorithm trading and market microstructure literature. We define two well-known impact functions: exponential and power law functions, linked to the exponential decay function. The parameter $\alpha$ is a proxy for the inverse of the market liquidity:\\
\begin{itemize}
\item
Power law impact function:
\begin{eqnarray*}
f(\boldsymbol{\gamma})=\boldsymbol{\gamma}^\alpha , \quad \alpha>0
    \end{eqnarray*}
\item
Exponential impact function:
\begin{eqnarray*}
f(\boldsymbol{\gamma})=exp(\alpha \boldsymbol{\gamma}) ,\quad \alpha>0
    \end{eqnarray*}
\end{itemize}
 \cite{gatheral2010} examined the dynamic-arbitrage for different combinations of the market impact function and decay function to not admit price manipulation strategies. (\cite{alfonsi2010}, proposition 2) proved the convex and non-constant exponential price impact, leads to the strictly positive definite property, to avoid arbitrage implication, which is already stated in \cite{gatheral2010}.
  Nonlinear impact function that is not a related state of the market is questionable; however, above functions defined in an illiquid market depend on the state of market.\\

\newtheorem{mypropos}[thm]{Proposition}
\fixstatement{mypropos}
Hawkes process 

We model the dynamics of trade arrivals intensity with the Hawkes process as a point process. This model can capture irregular properties of the high frequency data like the strength of an incentive to generate the same event with parameter $\sigma$. With choosing the price impact function $f$ as a function of a trading rate $\gamma_t$ and the parameter $\kappa$ as the exponent of the decay of market impact, we can model an impact of liquidation of a large position on the trade arrival dynamics as a particular form of the Hawkes process with the following SDE:
\begin{eqnarray}
d \lambda_t = (f(\gamma_t)-\kappa \lambda_t)dt + \sigma dN_t
\label{sdepriceimpact}
\end{eqnarray}

In the following proposition, we model the impact of trading on the rate of arrivals based on the Hawkes process:
\begin{mypropos}
\textbf{Price Impact Model}\\
Let $f$ be a price impact function and a function of a trading rate $\gamma_t$. $\sigma$ and  $\kappa$ represent the magnitude of self-exciting and the exponent of the decay of market impact, respectively. The solution of the SDE \label{sdepriceimpact} is expressed by:
\begin{eqnarray}
 \lambda_t = \int_0^t(f(\gamma_s)\Gamma(t-s))ds + \sigma \int_0^t e^{-\kappa(t-s)}dN_s
\end{eqnarray}
where the function $\Gamma$ is the decay of impact,  \\
\end{mypropos}
(Proof is given in the appendix)
\proofatend
 Following SDE represents the impact of trading on the dynamics of the rate of orders' arrival:
\begin{eqnarray}
d \lambda_t = (f(\gamma_t)-\kappa \lambda_t)dt + \sigma dN_t
\end{eqnarray}
In order to prove, we can move the first term of SDE to the left side, then multiply it by $e^{\kappa t}$ (see : \cite{norberg2004vasivcek}) or alternatively we can define an initial guess for the solution of above SDE as follows:
\begin{eqnarray}
\label{impactintensity}
 \lambda_t = \lambda_0 + \int_0^t(f(\gamma_s)\Gamma(t-s))ds + \sigma \int_0^t e^{-\kappa(t-s)}dN_s
\end{eqnarray}

Verify by It\^o's lemma on $e^{\kappa t} \lambda_t$
\begin{eqnarray}
 e^{\kappa t} \lambda_t  &=& e^{\kappa t}  \int_0^t(f(\tau_s)e^{-\kappa(t-s)})ds + e^{\kappa t} \sigma \int_0^t e^{-\kappa(t-s)}dN_s\\
            &=&  \int_0^t(f(\gamma_s)e^{\kappa s})ds + \sigma \int_0^t e^{\kappa s}dN_s\\
 \kappa e^{\kappa t} \lambda_t dt+e^{\kappa t}d\lambda_t &=&  (f(\gamma_t)e^{\kappa t})dt + \sigma  e^{\kappa t}dN_t\\
 \kappa \lambda_t dt+d\lambda_t &=&  (f(\gamma_t))dt + \sigma dN_t\\
   d\lambda_t &=&  (f(\gamma_t)-\kappa \lambda_t)dt + \sigma dN_t\\
\end{eqnarray}
\endproofatend

\begin{lem}
The impact stochastic intensity (equation: \eqref{impactintensity}) is a general function of price impact.  It can then measure a instantaneous price impact in the short term, and a permanent price impact on the long run.
\end{lem}
(Proof is given in the appendix)
\proofatend
of a buy order \eqref{impactintensity}. 
\begin{eqnarray*}
\lim_{t \to \infty} D(t) &=& \lim_{t \to \infty} \int_0^t(f(\gamma_s)\Gamma(t-s))ds\\
     &=& \lim_{t \to \infty} \int_0^t(f(\gamma_s)e^{-\kappa(t-s)})ds\\
     &=& \lim_{t \to \infty} e^{-\kappa t} \int_0^t(f(\gamma_s)e^{\kappa s})ds\\
     &=& \lim_{t \to \infty} \frac{\int_0^t(f(\gamma_s)e^{\kappa s})ds}{e^{\kappa t}}  \\
   (\mbox{apply l'H\^opital's rule})  &=& \lim_{t \to \infty} \frac{(f(\gamma_t)e^{\kappa t})}{\kappa e^{\kappa t}}  \\
   &=& \lim_{t \to \infty} \frac{exp(\alpha \gamma_t)}{\kappa}  \\
   &=& \lim_{t \to \infty} \frac{1+\alpha \gamma_t +O(\alpha \gamma_t)}{\kappa} \approx \frac{1+\alpha \gamma_T}{\kappa} = \lambda_{\infty} \\
    \end{eqnarray*}
We defined the liquidation problem as a finite time investing problem on a limited time horizon $T$. $\lambda_{\infty}$ represents the long run trading impact on the intensity of order arrivals rate, we can think of it as a permanent price impact as "base" intensity part of stochastic intensity. It is a linear function of trading rate to avoid dynamic arbitrage \cite{gatheral2010}. We can express the permanently effected stochastic intensity by:

\begin{eqnarray}
 \lambda_t^{Perm} = \lambda_{\infty} + \sigma \int_0^t e^{-\kappa(t-s)}dN_s
\end{eqnarray}

Instantaneous market impacts can be measured from the small interval of trading, and the difference between the pre-trade and post-trade price movements:

\begin{eqnarray*}
\lim_{\varepsilon \to 0} D(t) &=& \lim_{\varepsilon \to 0} \int_t^{t+\varepsilon}(f(\gamma_s)\Gamma(t-s))ds\\
     &=& \lim_{\varepsilon \to 0} \int_t^{t+\varepsilon}(exp(\alpha \gamma_s)e^{-\kappa(t-s)})ds \approx exp(-\kappa t + \alpha \gamma_{\varepsilon}) = \lambda_{\varepsilon}\\          
    \end{eqnarray*}

We can simply define the instantaneously affected stochastic intensity by:
 \begin{eqnarray}
 \lambda_t^{Inst} = \lambda_{\varepsilon} + \sigma \int_0^t e^{-\kappa(t-s)}dN_s
\end{eqnarray}   
\endproofatend

\begin{assumption}
\textbf{Continuity and concavity of the value function}\\
We assume that the value function $V$ is a continuous and bounded function. It is also strictly concave in $q$, increasing in $t$ and non-negative. The differentiability condition of the value function is not necessary to be satisfied.
\end{assumption}

\section{Solving Model by Discrete-Time MDP}
\label{Solving_model}
The liquidation problem \ref{mainproblem} is a continuous-time stochastic control problem.  \cite{bayraktar2014liquidation} formulated this problem as a Hamilton--Jacobi--Bellman equation, and provided viscosity solutions as closed-form solutions associated with the limit order book model and its depth function. The classical stochastic control approach solves nonlinear partial differential equations, and it is necessary the differentiability of the value function to be satisfied. In contrast, a discrete-time Markov Decision Processes (MDP) approach provides a set of optimal policies, condition on the differentiability of the value function. \cite{bauerle2009mdp} used this approach and by applying dynamic programming principles, proved the existence and uniqueness of the solution.\\

This liquidation problem is a mathematical abstraction of real problems in which an investor should make a decision on several stopping times to gain certain revenue at each stage. The investor has a finite period to liquidate a position, and maximize the total revenue at the end of the period. Therefore, in this problem with a finite number of sub-periods, a mapping function should be applied to compute optimal policies through the limited number of steps of the dynamic programming algorithm.\\

She must find an equilibrium to minimize the cost of the present extensive trading against the future abandon risk where the overall cost is not predictable. This problem can be formulated as a deterministic or stochastic optimal control problem with Markov or semi-Markov decision property under different setups.\\

\cite{dimitri1996stochastic} distinguished between the stochastic optimal control problems from its deterministic form regarding available information. In a deterministic optimal control problem,  we can specify a set of states and policy as a control process in advance. Thereby, a succeeding state is the function of the present state and its control process. On the other hand, in the stochastic control problem, controlling the succeeding state of the system leads to evaluate unforeseen states; therefore control variables that are no longer appropriate or have ceased to exist. In this paper, we use a Piecewise Deterministic Markov Decision Model (PDMD) to decompose the liquidation (problem: \ref{mainproblem}) as a continuous-time stochastic control problem into discrete-time problems.\\

\subsection{Solution by PDMP}
 Piecewise Deterministic Markov Process (PDMP), introduced by \cite{davis1984piecewise}, is now largely applied in various areas such as natural science, engineering, optimal control, and finance. The PDMP is a member of the c\'adl\'ag Markov Process family; it is a non-diffusion stochastic dynamic model, with a deterministic motion that is punctuated by a random jump process.\\

\begin{mydef}
\textbf{Piecewise Deterministic Markov Process}\\
A piecewise-deterministic Markov process (PDMP) is a c\'adl\'ag Markov process with deterministic emotion controlled by the random jump at jump time.
\end{mydef}

This process is characterized by three measurement quantities. The first feature is the transition measure $k$, which selects the post-jump location. The other quantities are deterministic flow motion $\phi$  between jumps and the intensity of the random jump  $\Lambda_t$, defined as Borel measures on the Borel sets of the state space $E$, and the control action space $A$.\\

In the state space $E$, set $(t,x)$ donates the desired process value at the jump time point $t$. In an embedded Markov chain as a discrete-time Markov chain, the state of the PDMP process can be defined as a set of components of the continuous trajectory $Z_k=(T_k,X_{T_k})_{k= 1, \cdots ,n}$ where $T_k$ is an increasing sequence of the jump time component and $X_t$ and $Z_t$ are a jump location component and a post jump location component, respectively: ($Z_t = X_t$ if $t=t_k$).\\

This process starts at the state $x_t$, and jumps with the Poisson rate process $\Lambda_t$ (fixed or time dependent) to the next state or hits the boundary of the state space. Stochastic Kernel $K(.| x_t , a_t) $ by measuring the transmission probability selects the next location of the jump given the current information on state and action. Each Markovian policy is a function of the jump time component ($T_k$) and the post jump location $(Z_{T_k})$ with the following condition:
\begin{equation}
  Z_t = \left\{
      \begin{aligned}
       \phi(X_t), \quad \mbox{for} \quad t < T,\\
       X_t, \quad \mbox{for} \quad t= T.\\
      \end{aligned}
    \right.
\end{equation}

\begin{figure}[h]
\begin{center}
   \includegraphics[scale=.30]{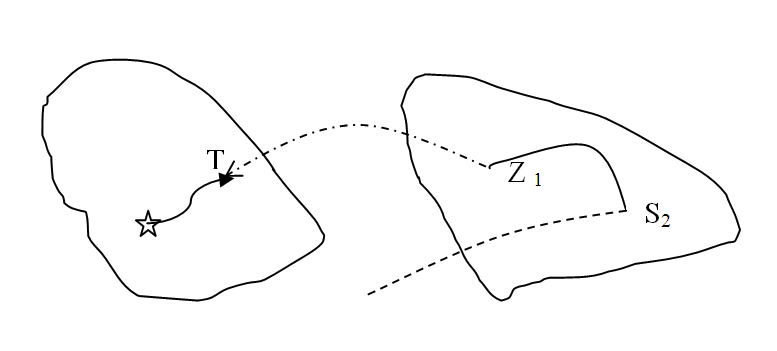} 
    \caption{Iterative Procedure of PDMP}
   \label{figPDMP} 
\end{center}
\end{figure}
Figure \ref{figPDMP} shows the iterative procedure of the PDMP: starting point of the PDMP is $X_0= Z_0$, then $X_0$ follows the flow $\phi$ until $T_1$ to determine first jump location  $X_1= \phi(X_0)$. The stochastic Kernel $K(.| \phi(X_{1}) ,.) $ selects the next location of the post-jump $Z_1= K(.| \phi(X_{1}) ,.)$. Latter, similar to the first jump, $X_t$ follows the flow $\phi$ until $T_2$ so $X_2= \phi(Z_1)$. In next step is the selection of a location of the post-jump $Z_2= K(.| \phi(Z_{1}) ,.)$ by using the stochastic Kernel $K(.| \phi(Z_{1}) ,.) $. This iterative procedure will be continued until it hits a boundary of the state space.\\

As mentioned earlier , the process of the illiquid asset in a finite discrete-time horizon $ \tau = \{ 0<T_1 \dots < T_m <n \}$ is:
    \begin{eqnarray*}
S_t=S^0 exp(\mu_t dt+ \sigma_t w_t)) \quad t \in \tau
    \end{eqnarray*}
and at time of jump (if $t \in [T_i,T_{i+1})$) the price process is:
\begin{eqnarray*}
S_{T_i}-S_{T_{i+1}}=S_{T_i} J_i
\end{eqnarray*}
where $J_i$ is an independent and identically distributed random variable. The price process $S_t$ is a so-called  piecewise-deterministic Markov process (PDMP).
\subsubsection{Markov Process Components}
General speaking, Markov process models, which are not stationary in our setup, include a set of the following terms:\\
 \begin{itemize}
  \item
Let $A$ be an action space includes the action $a_t$ which denotes the quantity of shares to be liquidated at time $t$. 
\begin{eqnarray*}
a_t =\gamma_t \bigtriangleup\\
 t \in[0, \cdots , T]
\end{eqnarray*}
where $\gamma_t$ is an admissible strategy which satisfies the condition $\gamma_t \leq \frac{q}{\bigtriangleup}$ to have a non-negative inventory. 
\item
With a given action set $A$, the set $E$ is defined as a state space, that contains the state $x$ of inventory after applying action $a$. The process state $X_t$ denotes the amount of not liquidated shares at jump time $t$.   
\begin{eqnarray*}
X_t&=&Q^0 - \int_0^t q_s dNs\\
   &=&Q^0 - \int_0^t \gamma_s \bigtriangleup dNs\\
\end{eqnarray*}
\item
Let $K$ be a stochastic transition kernel from $E \times A$, as a set of all state-action, to a set of states $E$. We measure the probability of the next state and action with transition kernel $K(.|x_t,a_t)$ based on the transition law at Markov decision model.
\item
The reward function $R$ represents the expected gain as a result of applying the strategy $\gamma$, of each state at the jump point time $t$:
\begin{eqnarray}
 R (t,x,\gamma)&=& \int^t_0 \gamma_s \bigtriangleup S_s dNs = \int^t_0 \gamma_s \bigtriangleup S_s \lambda_s ds\\ \label{sifunction}
               &=& \int^t_0 \mathcal{U}^{\boldsymbol{\gamma}}(X_s) dN_s
\end{eqnarray}
where the function $\mathcal{U}: (0,\infty) \rightarrow \mathbf{R_+}$ is an increasing concave function, which represents the preference over a set of limit orders or satisfaction of the trader from offers in the market and $X_s$ is the process state $X_t$ represents the amount of not liquidated shares at jump time $t$.  
\item
 The deterministic flow motion $\phi$ measures the movement of the inventory of investor between two jumps with a given the strategy $\eta$:
 \begin{eqnarray}
  \phi^\eta(X_t)= \int^t_0 \eta_s \bigtriangleup dNs 
\end{eqnarray}
  \end{itemize}
  
To be more precise on how we can solve the liquidation problem with PDMD, we explain the formal expression of deterministic optimal control problems, which is well documented in \cite{dimitri1996stochastic}. In a formal deterministic optimal control problem, $x_i$ presents the state of the system at stage $i$ and  function $c_i$ is a corresponding control at that stage. System equation $x_{i+1}=f(x_i,c_i)$ is the generating function of next state $x_{i+1}$ from current state $x_i$ and its control $c_i$. Function $g(x_i,c_i(x_i))$ is the rewarding function of state $i$ a associate with function $c_i$.\\
 The total expected revenue after $N$ decision steps is defined by:
\begin{eqnarray}
J^{\mathcal{C}}(x_0) = \mathbb{E}_{t,x}[ \sum^N_{i=0} g^{\mathcal{C}}(x_i,c_i(x_i))]
\end{eqnarray}

The set $\mathcal{C}$ contains all control functions $(c_i)_ {i=\{0,\cdots,N\}}$, i.e. the expected total revenue is set of states and corresponding a sequence of Markovian decision controls.\\

Let $\Pi=\{\pi_1, \pi_2,\cdots,\pi_N\}$ be the sequence of all Markovian decision controls $\pi_i$. These decision controls are corresponding to Markovian policy for predictable admissible process $\boldsymbol{\gamma}$ defined in the set $\Psi$. Each Markovian policy is a function of the jump time component ($T_k$) and the post jump location $(Z_{T_k})$. \\
In the each stage, the reward obtained as a result of sequential Markovian decisions after the jump by applying the strategy $\pi_i$ is:
   
  \begin{eqnarray}
 R (Z_i,\pi_i(Z_i))= \int^t_0 \mathcal{U}^\pi(\phi_s( X_i))dN_s
\end{eqnarray}
The total expected revenue after $N$ steps is defined by:
\begin{eqnarray}
\Psi^\pi(t,x) = \mathbb{E}_{t,x}[ \sum^N_{i=0} R(Z_i,\pi_i(Z_i))]
\end{eqnarray}

In the following theory, we explain how the above equation is mathematically equivalent to our main problem: (\ref{mainproblem}). 


\begin{thm}
\label{optimaltheorm}
Suppose policy set $\Pi=\{\pi_1, \pi_2,\cdots,\pi_N\}$ is a Markovian policy set and $Z=\{Z_1,Z_2,\cdots,Z_n \}$ is a set of post jump location of PMDP and $R$ is the reward function (equation: \ref{sifunction}). Then the value function is the expected reward of the PDMP under the Markovian policy $\Pi$ at time point $t$ and in the state $x$:
  \begin{eqnarray*}
V(t,q) = \sup_{\pi\in \Pi} \Psi^\pi(t,x) , \quad  t \in \tau , \quad q \in [0,Q^0]
  \end{eqnarray*}

where
  \begin{eqnarray*}
\Psi^\pi(t,x) = \mathbb{E}_{t,x}[ \sum^N_{i=0} R(Z_i,\pi_i(Z_i))].
\label{pdmdfunc}
 \end{eqnarray*}
\end{thm}
(Proof is given in the appendix)
\proofatend
 By using the information on the jump location from  $Z=(Z_1,Z_2,\cdots,Z_n)$ as a set of post jump of PMDP , we have:
 \begin{eqnarray*}
   V^\pi(t,q)&=&   \mathbb{E}_{t,q}[\int_{0}^{T\prime }e^{-r t}  \bigtriangleup \gamma_t s_t  \hat{\lambda_t} dt] \quad (t \in \tau) \\   
  (\mbox{refer to equation:} \ref{sifunction}) &=&   \mathbb{E}_{t,x}[\int_{0}^{T\prime } \mathcal{U}^\pi(X_{T\prime}) dN_t]\\
                                               &=&  \mathbb{E}_{t,x}[\sum_{i=0}^{N}[\int_{T_i}^{T_{i+1}\wedge T\prime}\mathcal{U}^\pi(\phi(X_{T_i})) dN_t ]\\
   (Z_i  \mbox{ define as } Z_i=[T_i,X_{T_i}])  &=&   \mathbb{E}_{t,x}[\sum_{i=0}^{N} \mathbb{E}_t[\int_{T_i}^{T_{i+1}\wedge T\prime}\mathcal{U}^\pi(\phi(X_{T_i})) dN_t |Z_i]]\\
                                                  &=&   \mathbb{E}_{t,x}[\sum_{i=0}^{N}  R(Z_i,\pi_i(Z_i)) ]\\
                                                  &=&   \Psi^\pi(t,x)
         \end{eqnarray*}
We define a set $\Pi=\{\pi_1, \pi_2,\cdots,\pi_N\}$ as a sequence of all Markovian decision controls $\pi_i$ corresponding to the Markovian policy for the predictable admissible process $\boldsymbol{\gamma}$ included in the set $\Psi$. We can then decompose this optimal control problem into the piecewise-deterministic Markov process:         
  \begin{eqnarray*}
V(t,q) = \sup_{\pi \in \Pi} \Psi(t,x)
  \end{eqnarray*}               
\endproofatend 

\subsection{Uniqueness of solution}

We started to model the optimal liquidation problem as a stochastic control model (equation \ref{maincerterifunction}) and used the piecewise-deterministic Markov process to find an equivalent deterministic model for this problem. We also proved this problem can be constructed as a summation of the sequence of state processes and corresponding control processes in the set $\Pi$. The formulation has been given in the equation \ref{mainformalprobelm} and is more consistent with dynamic programming principle.\\
\cite{dimitri1996stochastic} defined the universal measurable mapping $T$ to map equation \ref{maincerterifunction} from \ref{sifunction} as follows:

\begin{eqnarray}
\mathcal{T}^{\pi}(\Psi)(x) = H [x,\pi,\Psi]
\end{eqnarray}
 Therefore the operator $\mathcal{T}^{\pi_n}$ can be decomposed to $\mathcal{T}^{\pi_0}\cdot \mathcal{T}^{\pi_1}\cdot \mathcal{T}^{\pi_2} \cdots \mathcal{T}^{\pi_{n-1}}$ . \\
From above definitions, we have:
 \begin{eqnarray*}
{\Psi}^{\pi_{n}} &=& \mathcal{T}^{\pi_n}(r)\\
            &=& \mathcal{T}^{\pi_0} \mathcal{T}^{\pi_1} \cdots \mathcal{T}^{\pi_{n-k}} {\Psi}^{\pi_{n-k}}
\end{eqnarray*}
The mapping $\mathcal{T}^{\pi}$ is an universal measurable mapping, let $\mathcal{T}^{\pi_0}=0$ and $k \leq n$,  (see: \cite{dimitri1996stochastic}(chapter 1) we have:
\begin{eqnarray*}
{\Psi}^{\pi_{k}} = \mathcal{T}^{\pi_0} \mathcal{T}^{\pi_1} \cdots \mathcal{T}^{\pi_{k-1}} {\Psi}^{\pi_{0}}
 \end{eqnarray*}
In the following theorem, we prove the uniqueness of the solution by applying a piecewise-deterministic Markov process model.

\begin{thm}Uniqueness of the solution\\
Let $V(t,q)$ be a revenue function and the optimal solution of the liquidation problem is a concave and decreasing function in $q$ and increasing in $t$. Then the solution of the liquidation problem by applying a PMDP model is converging to a unique solution.
\end{thm}
(Proof is given in the appendix)
\proofatend
In the theorem \ref{optimaltheorm}, we have shown that 
\begin{eqnarray*}
V(t,q) &=& \sup_{\boldsymbol{\gamma}} H^{\boldsymbol{\gamma}}(t,q)\\
           &=& \sup_{\pi \in \Pi}\Psi^{\pi}(t,x) \\
\end{eqnarray*}
with defining the sequence of $\Pi = \{\pi_1, \pi_2,\cdots,\pi_n\}$ as set of Markovian policies (\cite{dimitri1996stochastic}), we have:
\begin{eqnarray*}
\Psi^{\pi_n} &=&  \lim_{n \to +\infty} \mathcal{T}^{\pi_0}\cdot \mathcal{T}^{\pi_1} \cdot \mathcal{T}^{\pi_2} \cdots \mathcal{T}^{\pi_{n-1}} \Psi^{\pi_{0}}(x_0)\\
               &=& \lim_{ n \to +\infty} (\mathcal{T}^{\pi_n})^n (\Psi^{\pi_0}(x_0))\\
             &=& \mathcal{T}^n (\Psi^{\pi_{n}}(x_0))\\
 \end{eqnarray*}
 Therefore under the same condition, the optimal solution is defined as:               
         \begin{eqnarray*}
             \Psi(x) =\sup_{\pi \in \Pi}(\Psi^{\pi}(x))\\
               \end{eqnarray*}      
   Equally
                   \begin{eqnarray*}
               \Psi(x) =\mathcal{T}(\Psi(x))\\
               \end{eqnarray*}                  
which is a subset of the Banach space, so we can apply the Banach fixed point theorem and show that $V$ is an unique fixed point of the operator $\mathcal{T}$ on the set $\Pi$.
\endproofatend

\section{Numerical method and Simulation}
\label{Numerical}
In this section, we apply a simulation method to assist the performance of our model under various market microstructures' characteristics. The trader will decide to take a number of offers in the LOB at given prices. The optimum trading rate is dependent on the dynamics of orders' arrival as well as time to maturity. We approximate the value function with a quantization method.

\subsection{Approximation of the value function}

As we explained before, the solution of the value function $V$ of the optimal liquidation problem is obtained by the summing rewards of sequential Markovian decisions with corresponding the Markovian policies $\pi$ and a set of post jump process of the PMDP:   

\begin{eqnarray*}
       V(t,Q^0)  &=& \sup_{\boldsymbol{\gamma} \in \mathcal{A}}\mathbb{E}_{t,q}[V(t+1,Q^0-q)+ h(t,q)]\\
                 &=& \sup_{\boldsymbol{\gamma} \in \mathcal{A}}\mathbb{E}_{t,q}[\sum_{i=0}^{\infty} R(Z_i,\pi_i(Z_i))\textbf{1}_{T_i<T}+ h(t,q)\textbf{1}_{T_i \geq T}]
                 \end{eqnarray*}
We approximate the value function $V$ with the function $\hat{V}$, such that $|V-\hat{V}|_{Lp}$ is minimized for the $Lp$ norm. To approximate the continuous state space by a discrete space, we use a technique that is called Quantization method. \cite{bally2003quantization} and \cite{bally2005quantization} developed quantization methods to compute the approximation of a value function of the optimal stochastic control. \cite{de2010numerical} explained the implication of the numerical solution to PMDP, such that transmission function cannot be computed explicitly from local characteristics of PMDP.  \cite{de2010numerical} expressed a numerical solution for Embedded Markov chain, while the only source of randomness is a set of the post jump process $(T_n, Z_n)$. By quantization of $Z_n=(X_{T_n}; T_n)$, we can transfer the conditional expectations into finite sums, and least upper bound of the value function ($\sup$) into its maximum value ($\max$) in discretized space of $[0,T]$. We define $\hat{V}$ as an \textit{approximation of the value function} as follows:        
\begin{eqnarray*}
       \hat{V}(t,Q^0)  &=& \max_{\boldsymbol{\gamma} \in \mathcal{A}}\mathbb{E}_{t,x}[\sum_{i=0}^{\infty} R(\hat{Z}_i,\pi_i(\hat{Z}_i))\textbf{1}_{\hat{T}_i<T}+ h(t,q)\textbf{1}_{\hat{T}_i \geq T}]
                 \end{eqnarray*}

\cite{de2010numerical} estimated the error and the convergence rate of the approximated value function with Lipschitz assumption of local characteristics of PMDP, and showed it is bounded by the constant rate of quantization error $Qe$.
\begin{eqnarray}
|V(t,Q^0) - \hat{V}(t,Q^0)| \leq Qe
\end{eqnarray}

\subsection{Simulation}
\begin{figure}[h]
\begin{center}
   \includegraphics[scale=.35]{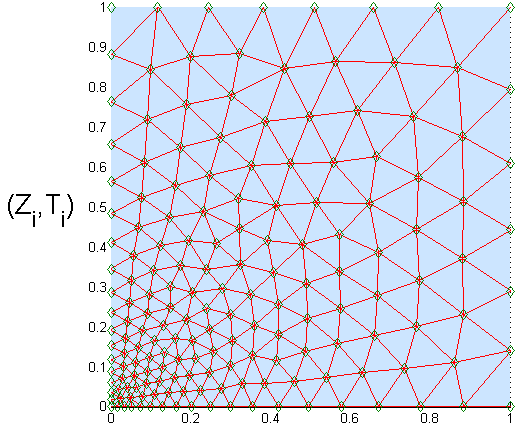} 
      \caption{Quantization of state space $(Z_i,T_i)$}
  \label{picmesh}    
\end{center}
\end{figure}

Conditional expectation of value function can be computed with some numerical methods in finite dimensional space, such as regression method \cite{longstaff2001valuing} or on Malliavin calculus (as in \cite{cont2010change}). \cite{bally2003quantization} proposed a quantization method to approximate the state space of problem; from each time step $T_k$, a sate function $\hat{Z}$ can be projected to the grid $\Upsilon_k := \{\hat{z}_{T_k^i} \}_{1 \leq i \leq N}$ (see: figure \ref{picmesh})
\begin{eqnarray}
\hat{Z}_{T_k}=\sum_{1 \leq i \leq N} \hat{z}_{T_k^i} \mathbf{1}_{T_k^i \in \mathcal{B}^k_i}
\end{eqnarray}
where $\mathcal{B}^k_i$ is a Borel partition of $\mathbb{R}^d$ (see: \cite{bally2003quantization}).\\

As previously stated, at each stage, the reward obtained for each stage of sequential Markovian decision after a jump by applying strategy $\pi_i$ is:
   
  \begin{eqnarray}
\hat{V}_k= R (\hat{Z}_{T_k},\pi(\hat{Z}_{T_k}))
\end{eqnarray}
By applying dynamic programming principle for $n$ fixed grids $\Upsilon_{0 \leq k \leq n}$, $\hat{V}_k $ satisfies the backward dynamic programming condition: 
  \begin{eqnarray}
\hat{V}_n= R (\hat{Z}_{T_n},\pi(\hat{Z}_{T_n}))
\end{eqnarray}

  \begin{eqnarray}
\hat{V}_{k}= \max( R (\hat{Z}_{T_k},\pi(\hat{Z}_{T_k}),\mathbb{E}(\hat{V}_{k+1}|\hat{Z}_{T_k}))
\end{eqnarray}

We have used a particular SDE form of the Hawkes process of the intensity of the rate of order arrivals to express the impact of order execution on the market:
\begin{eqnarray}
d \lambda_t = (f(\gamma_t)-\kappa \lambda_t)dt + \sigma dN_t
\end{eqnarray}

where $f(\gamma_t)$ is a function of trade rate $\gamma_t$, $\sigma$ explained the strength of the incentive to generate the same event , and $\kappa$ is the exponent of the decay of market impact.\\

The Hawkes process can capture both exogenous impacts and endogenous influence of past events to measure the probability of occurrences of events. This intensity is based on modeling with ($\sigma > 0$ ) to capture the contribution of past events to amplify the chance of an occurrence of the same type of events (Self-exciting property of the Hawkes process). By using this endogenous mechanism, (while $\sigma < 0$), the model can explain the significant aspect of the strategic function of market participants known as a "market manipulation" (see:, e.g., \cite{cartea2014buy}). It is often used by traders to submit or cancel orders strategically to detect hidden liquidity or to manipulate markets.\\

If trading activity with reducing possibility occurrence  has an adverse influence on the arrival of orders, the jump process has an adverse impact on its intensity and makes imbalance in supply and demand of the market exponentially. This damping factor can be measured with Hawkes model the magnitude of self-exciting and the strength of the incentive while it is negative (self-damping property).

\subsection{Result of Simulation }

In this part, we present the numerical solution of the optimal liquidation problem. To study characteristics of the value function and the level of inventory associated with the control variable $\gamma$ as the rate of trading, we compute some numerical examples using different scenarios from empirical studies. Our simulation is an abstract of real problems. We consider a trader who wants to liquidate $Q^0$ shares of a risky asset within a short time and fixed time horizon $T$.  In an illiquid market, she expects a longer time to liquidate the whole position. Her goal is to minimize the implicit and explicit cost of trading and keep the low-level inventory by controlling the trading rate.\\

We implement our model in the discrete state space. We choose time steps small enough to increase the chance to catch orders and larger than the usual tick time to make sure that quotes are not outside of the market bid-ask spreads. We assess the performance of our model by quantizing the value function at a fixed position in the space and time of the mesh refinement.\\

To illustrate in more detail how our model behaves, we study the inventory level associated with the optimal trading rate control for both types of price impact Hawkes models: self-damping and self-exciting properties. Concerning our simulation scenarios, we choose values of the parameters of the model: (a) the parameter $\sigma$ as the magnitude of self-exciting, (b) the parameter $\kappa$ as the exponent of the decay of market impact and (c) the size of the bid-ask spread is used as a proxy to measure the illiquidity. These parameters can be estimated from the real market data.\\

\begin{figure}[h]
\begin{center}
   \includegraphics[scale=.65]{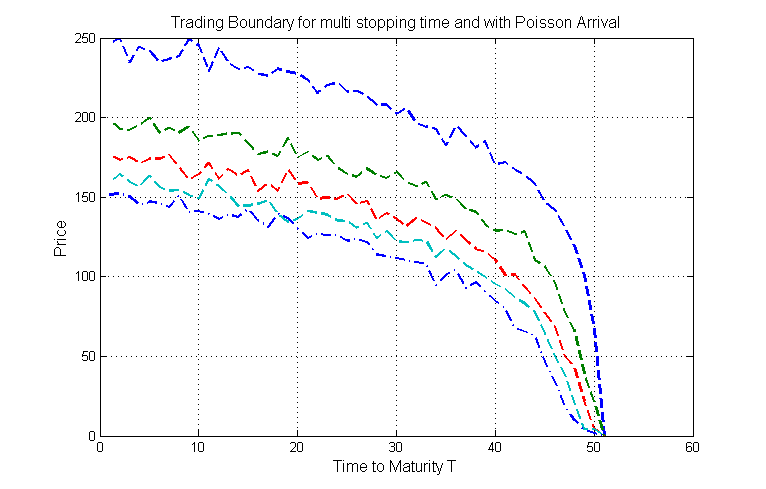} 
      \caption{Trading boundary condition for multi-Stopping time and with Poisson arrival}
 \label{tradingBoundry}
\end{center}
\end{figure}

Plotting the trading boundaries (figure \ref{tradingBoundry}) shows that the optimal trading level depends on the price process, the remainder of the inventory, and time to maturity.\\  

Due to market's conditions and asset's characteristics, an agent with the higher degree of risk aversion, cares more about the execution risk  and price fluctuations. She starts the trading with available orders at a deeper level of the limit order book to avoid the risk execution and lack of offers in the future. She splits the original order into smaller slices to mitigate price impacts.\\ 
\begin{figure}[h]
\begin{center}
\subfloat[]{\includegraphics[scale=.35]{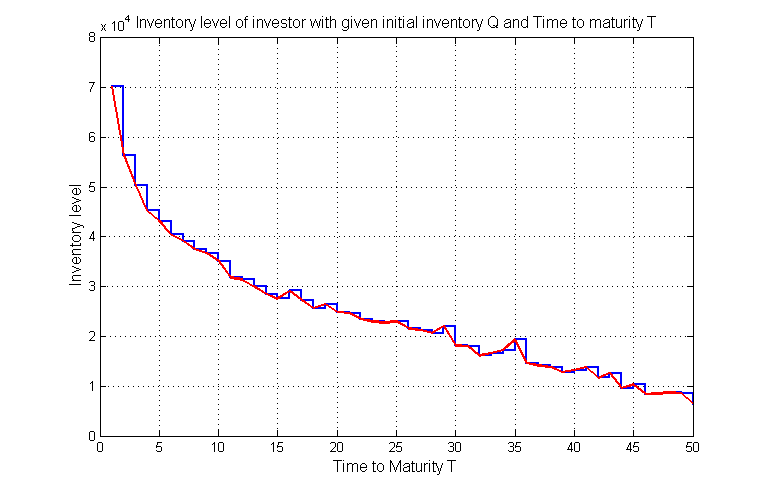}} 
\subfloat[]{\includegraphics[scale=.35]{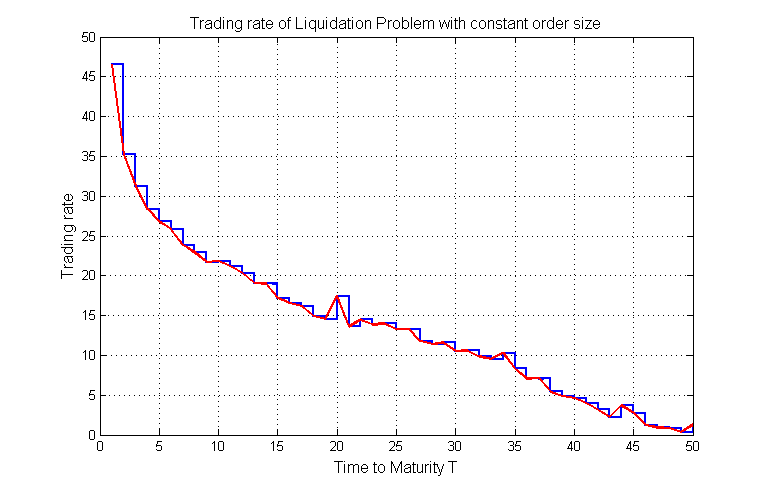}}
\caption{(a): Inventory level of a trader with given the initial inventory $Q^0$ and time to maturity T, (b): Trading rate of liquidation problem with constant order size $\bigtriangleup$ } 
\end{center} 
\label{rateInve}
\end{figure}

\begin{center}
\begin{table}[h]
\centering
\begin{tabular}{cclcccccc}
 \hline\\Panel I&&&&&Quantile&\\
 Self-damping&Revenue&&10\% & 25\%        & 50\%            & 75 \%       & 100\% \\\hline
 $\kappa=0.6$    & \$50,067& Trad_rate    & 0.36    & 6.17     & 12.75    & 19.28        & 44.16    \\
  $\sigma=-0.6$   &             & Inven_level & 6205.45 & 14980.97 & 22618.06 & 31680.03 & 70000 \\\hline
  $\kappa=0.2$   & \$50,352 & Trad_rate    & 0.27    & 5.69     & 11.92    & 19.09    & 44.59    \\
$\sigma=-0.1$     &             & Inven_level & 6479.85 & 15814.89 & 23071.59 & 31534.80 &  70000 \\\hline
 $\kappa=0.6$    & \$50,000 & Trad_rate    & 0.13    & 5.53     & 11.36    & 21.00    &  40.38    \\
  $\sigma=-0.1$   &             & Inven_level & 4159.49 & 14490.57 & 20724.59 & 30499.13 & 70000 \\\hline\\  Panel II&&&&&Quantile&\\ Self-exciting&Revenue&& 10\% & 25     \% & 50\%            & 75\%        & 100\% \\\hline
 $\kappa=0.6$    & \$50,779 & Trad_rate    & 0.59    & 6.17     & 12.54    & 19.27    &  43.20    \\
   $\sigma=0.1$  &             & Inven_level & 3229.40 & 14751.81 & 22830.39 & 30507.01 & 70000 \\\hline
   $\kappa=0.2$  & \$62,539 & Trad_rate    & 0.03    & 5.74     & 13.16    & 22.15     & 48.74    \\
   $\sigma=0.1$  &             & Inven_level & 1593.42 & 11177.87 & 18155.91 & 31810.18 & 70000 \\\hline
$\kappa=0.6$     & \$94,691 & Trad_rate    & 0.34    & 9.68     & 16.12    & 26.71    &  72.67    \\
 $\sigma=0.6$    &             & Inven_level & 2153.13 & 7580.03  & 19473.28 & 36405.45 &  70000\\
     \hline
\end{tabular}
\caption{Summary of the result of simulation the level of inventory and its corresponding optimal trading rate under different market conditions}
\label{mytable1}
\end{table}
\end{center}
Table \ref{mytable1} summarizes results of simulations by our model, including the level of inventory and its corresponding optimal trading rate for different scenarios of implementations of strategies. In the case of coming of not favorite offers, the algorithm reduces the speed of trading (panel I: self-damping property) and waited for a longer time to find better matching counterparties. In unstable market conditions, are indicated with the higher level of self-damping, the trader should pay for final inventory to liquidate the whole position of initial shares. At this point, the best strategy is to accept offers in the limit order book to avoid to never face severity penalties at the end of period. If estimated parameters of markets might show that the higher chance of same types of orders' occurrences (panel II: self-exciting property), algorithm reduces the trading rate in the hope of getting better offers. The second column shows the related entries of the value function as a result of the implicit and explicit cost of trading. The results demonstrate that the second type of the market characteristics with the higher level of self-exciting property can be more profitable.\\

\begin{figure}[h]
\begin{center}
   \includegraphics[scale=.55]{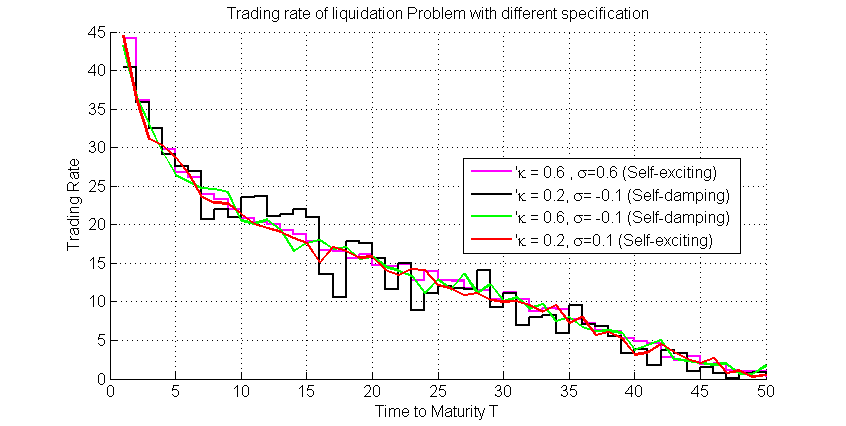} 
         \caption{Trading rate of liquidation Problem with different specification}
  \label{tradRatscenar} 
\end{center}
\end{figure}
Having a look on the graphs (figure \ref{tradRatscenar})  showing the optimal trading rate of different scenarios, 
one can conclude that there are more fluctuations in trading if the probability of arrival new offers is reduced.\\

Our numerical results also show that the proportion of the gain of the liquidation model considerably depends  on the specification of the price impact function. It indicates that market conditions have an effect on the final inventory level and causes a substantial dropping in the final wealth of the trader. An important aspect of the optimal strategies, which we have developed, is to take into account the execution risk in an illiquid market i.e. inability to liquid shares at the given time.\\

The main assumption of the majority of limit order models is to trade at best bid and ask prices (see: \cite{cao2008order}, \cite{battalio2014can}). We allow the trading procedure go to deeper into the limit order book to avoid not filling the order and face last minute inventory penalties.


\section{Discussion and Further works}
\label{summarizes}
In this paper, we proposed an analytical solution for the optimal liquidation problem with a dynamic approach and build numerical boundaries of multi-stopping problems in an illiquid market. We simulated the optimal splitting orders models according to the existing liquidity in the order book with different parameters and price impact models. We have used a Piecewise Deterministic Markov Decision Model (PDMD) to decompose the liquidation problem, as a continuous-time stochastic control problem, into discrete period problems and applied Markov decision rules to obtain the solution. We studied the uniqueness and existence of the optimal solution. We indicated that that the percentage gain of the liquidation model depends on the market conditions and specification of the price impact function.\\

In direct opposition to majority the limit order models for liquidating market which only trading at best bid and ask prices, our model allows the trading to go deeper into limit order book to avoid not filling of the order and face last minute inventory punishment.\\

We believe that an attractive extension of our work would be to study; \cite{cartea2014optimal} discussed sophisticated models to trade at market order and post limit order; \cite{cartea2014algorithmic} found optimal combinations of market and limit orders with learning from market dynamics to trade in the direction of price fluctuations.\\

Modeling the orders' arrival  flow with a Poisson process is  quite robust approach. \cite{cartea2013robust} addressed several uncertainties about the arrival rate of orders, the risk of not filling with limit orders, and misspecification in the dynamics of the stock's midprice, with the robust portfolio optimization approach.  \cite{iyengar2005robust} proposed a robust formulation which is systematically alleviated the sensitivity of the Markovian policy on the uncertainty of transition probabilities. We would like to examine the robustness of our findings with relaxing some assumptions of the model.
\pagebreak
\section*{Reference}

\bibliographystyle{plainnat}
\bibliography{OrderBook}
\pagebreak
\section{Appendix}
\appendix
\printproofs


\end{document}